\documentstyle[12pt,epsfig]{article}
\newcommand{\lbar}{\bar{\Lambda}}

\newcommand{\beq}{\begin{equation}}
\newcommand{\eeq}{\end{equation}}
\newcommand{\bea}{\begin{eqnarray}}
\newcommand{\eea}{\end{eqnarray}}
\newcommand{\bdm}{\begin{displaymath}}
\newcommand{\edm}{\end{displaymath}}

\newcommand{\as}{\alpha_s}
\newcommand{\GeV}{\,\mbox{GeV}}

\newcommand\gsim{\mathop{\mbox{\vbox{\hbox{$>$} \vskip -9pt \hbox{$\sim$}
             \vskip -3pt  }}}}
\newcommand\lsim{\mathop{\mbox{\vbox{\hbox{$<$} \vskip -9pt \hbox{$\sim$}
             \vskip -3pt  }}}}

\newcommand{\msp}[1]{\mbox{\hspace*{#1mm}~}}

\begin{document}
\thispagestyle{empty}
\vspace*{-10mm}
 
\begin{flushright}
DFTT-11/05\\
Bicocca-FT-05-9\\
UND-HEP-05-BIG\hspace*{.08em}01\\

\vspace*{2mm}
\end{flushright}
\vspace*{25mm}
 
\boldmath
\begin{center}
{\LARGE{\bf
Hadronic mass and $q^2$ moments of charmless\\  \vspace*{3mm}semileptonic
  \boldmath $B$ decay  distributions 
}}
\vspace*{4mm}

\end{center}
\unboldmath
\smallskip
\begin{center}
{\Large{Paolo Gambino, Giovanni Ossola,}}  \\
\vspace{2mm}
{\sl INFN, Sez.\ di Torino and Dipartimento di Fisica Teorica, 
Universit\`a di Torino,\\   10125 Torino, Italy}  \vspace*{2.5mm}\\
{{\large  and}} \vspace*{3.0mm}
\\
{\Large{Nikolai~Uraltsev$^{*}$
}} \vspace*{2mm} \\
{\sl INFN, Sezione di Milano,  Milano, Italy}
\vspace*{18mm}

{\bf Abstract}\vspace*{-.9mm}\\
\end{center}
\noindent
We report OPE predictions for hadronic mass and $q^2$
moments in inclusive semileptonic 
$B$ decays without charm, taking into account 
experimental cuts on the charged lepton energy and on 
the hadronic invariant mass, and address the related theoretical uncertainty.

\setcounter{page}{0}
\vfill

~\hspace*{-12.5mm}\hrulefill \hspace*{-1.2mm} \\
\footnotesize{
\hspace*{-5mm}$^*$
On leave of absence from Department of Physics, University of Notre
Dame, Notre Dame, IN 46556, USA\\ 
\hspace*{-5pt}and from Petersburg Nuclear Physics
Institute, Gatchina, St.\,Petersburg  188300, Russia}
\normalsize

\newpage
\section{Introduction}

The precise measurement of the $V_{ub}$ element of the CKM quark mixing matrix 
from semileptonic $b\to u$ decays 
is one of the most important goals for the B-factories.
The high statistics accumulated by BaBar and Belle has recently 
made   a  measurement of the  
hadronic invariant mass distribution in inclusive  $B\to X_u \ell \bar\nu$ 
decays possible for the first time \cite{babarspectrum}.
The new generation of analyses is based on fully reconstructed events, which 
allows high discrimination between charmless events and charmed background,
even for hadronic invariant mass $M_X\gsim1.7$~GeV. 
After unfolding detector and selection effects, BaBar has been able to 
measure the  invariant mass distribution and two of its moments 
with promising accuracy. The measurements are possible even {\it without}
an upper cut on $M_X$, although it is clear that the relative error is 
smaller if one cuts at  $M_X^{\rm cut}$ 
close to the kinematic boundary for charm production 
(for instance, Ref.\cite{babarspectrum} adopts 1.86 GeV). The hope is that 
$M_X^{\rm cut}$ can be raised enough to suppress non-perturbative effects that 
cannot be accounted for by the local Operator Product Expansion (OPE), 
namely Fermi motion effects related to the $B$ meson distribution function(s),
without compromising the experimental accuracy.
Eventually, the measurement of the whole $M_X$ spectrum with
this new experimental technique could provide complementary information on the 
distribution function and possibly a very clean extraction of $V_{ub}$.

Once  accurate measurements of the moments of 
$B\to X_u \ell \bar\nu$ distributions are available, the 
first task is to verify 
their consistency with moments of $B\to X_c \ell \bar\nu$ and 
$B\to X_s \gamma$ in the OPE framework. Present analyses of semileptonic and
radiative  moments \cite{fits} show an impressive consistency of all the 
available data and the non-perturbative parameters they provide agree 
with independent theoretical and experimental inputs. They 
determine the $b$ quark mass within about 50 MeV and measure
the  expectation values of the dominant 
power suppressed operators with good accuracy.

In this paper we extend a previous analysis of $B\to X_c \ell \bar \nu$
moments in the kinetic scheme \cite{btoc} to the $B\to X_u \ell \bar \nu$ 
case. 
The moments in radiative decays in the kinetic scheme
have been studied in \cite{Benson:2004sg}.
Perturbative and non-perturbative corrections to the moments
of $B\to X_u\ell \bar \nu$ are given by  the $m_c\to 0$
limit of those to the  $B\to X_c \ell \bar \nu$ moments (see 
\cite{btoc,btocpert,dfn} and Refs.\ therein), but the upper cut 
on the hadronic invariant mass, $M_X^{\rm cut}$, requires a dedicated study.
We include all non-perturbative corrections through $O(1/m_b^3)$ 
\cite{1mb2,1mb3} and 
perturbative contributions through $O(\alpha_s^2 \beta_0)$ \cite{btocpert,dfn}.
We also investigate the range of $M_X^{\rm cut}$ for which the local OPE can 
be considered valid and give estimates of the residual theoretical 
uncertainty. A peculiarity of the $b\to u $ case is 
the presence of a  logarithmic divergence in the Wilson 
coefficient of the $1/m_b^3$ correction \cite{1mb3,bds}, related to the mixing 
between four-quark  and the Darwin operators. 
Fortunately, the hadronic moments are relatively insensitive to the ensuing 
uncertainty.
Finally, unlike the $b\to c$ case, 
all perturbative and non-perturbative corrections to the $b\to u$ case
can be expressed in terms of simple analytic formulas, 
if only a lower cut on the electron energy is used. 

We also consider moments of the $q^2$ distribution.
Experimentally, a measurement of the $q^2$ moments may soon 
become possible with only a lower cut on $E_\ell$. Provided $E_\ell^{\rm cut}$
is not higher than say 1.5~GeV, the local OPE prediction should be reliable.
That would be a very interesting measurement, as 
the higher $q^2$ moments could efficiently isolate the effect of 
the Weak Annihilation (WA) contributions which are concentrated 
at high $q^2$ values. From 
a practical point of view, as we will explain later on,
the $q^2$ moments can be calculated 
using the same building blocks as the invariant hadronic mass moments.

The paper is organized as follows: 
in Section 2 we briefly recall the formalism, define our notation, 
discuss the kinematics involved in the experimental cuts, and describe the way 
we perform the calculation. We also provide tables of reference values and 
approximate  expressions, and discuss $q^2$ moments.
In Section 3 we discuss the theoretical uncertainty of our results and
the effect of Fermi motion. 
Section 4 summarizes our main results. 
In Appendix A and B we report analytic formulas for the 
non-perturbative and perturbative corrections to the moments in 
the case of a cut on the lepton energy only.

\section{The calculation}
We consider the normalized integer moments of the squared invariant mass, 
\beq
\langle M_X^{2n}\rangle = \frac{\int d  M_X^2 \ M_X^{2n}\  d\Gamma/d M_X^2}
{\int d  M_X^2\  d\Gamma/d M_X^2}
\eeq
and introduce the notation
\beq
U_1= \langle M_X^{2}\rangle, \ \ \ \ \ \ \ \ \ U_{2,3}=\langle
\left( M_X^{2}-\langle M_X^{2}\rangle\right)^{2,3}\rangle
\label{Ui}
\eeq
for the first three central moments.
The physical hadronic invariant mass is related to parton level quantities by
\beq 
M_X^2=\bar \Lambda^2 +2m_b \bar \Lambda E_0 + m_b^2  s_0
\label{mx2}
\eeq
where $\lbar \equiv M_B - m_b$
($\bar\Lambda$ is defined here to include 
all power suppressed terms),  $E_0= 1-v\cdot q/m_b$, 
and $ s_0=(v-q/m_b)^2$, $q^\mu$ is 
the  four-momentum of the leptonic pair, and $v_\mu$ the four-velocity of the 
heavy meson. The moments of the parton level quantities $E_0$ and $s_0$  
and  
of their product are obtained in the local OPE and are expressed in terms of 
the heavy quark parameters; in particular, they do not depend on $M_B$ or, 
equivalently, on $\bar\Lambda$.
The building blocks in the calculation of the moments in 
Eq.~(\ref{Ui}) are therefore 
\bea
{\cal M}_{(i,j)}&=& \frac1{\Gamma_0}\int  d E_0\, d s_0 \,d E_\ell 
\ s_0^i \ \label{had_mom}
 E_0^j \ \frac{d^3 \Gamma}{d E_0 \,d s_0 \,d E_\ell} \\
\nonumber&=&
M_{(i,j)}+\frac{\alpha_s}{\pi} A_{(i,j)}^{(1)}+\frac{ \alpha^2_s\beta_0}
{\pi^2}A_{(i,j)}^{(2)}+ 
...\label{Mij}
\eea
where $\Gamma_0=G_F^2 m_b^5/(192\pi^3) $ is the {\it total} tree-level 
width, $\beta_0=11-2/3 \,n_f$ with $n_f=3$, 
and $M_{(i,j)}$ contains the tree-level contributions as well as 
non-perturbative corrections through $O(1/m_b^3)$.
We compute the non-perturbative corrections using \cite{1mb2,1mb3}, while
the perturbative corrections $A_{(i,j)}^{(1)}$ and $A_{(i,j)}^{(2)}$
are obtained (in the on-shell scheme) using the FORTRAN code
accompanying Ref.~\cite{btocpert} with a suitably small $m_c$ value.
The one-loop corrections $A_{(i,j)}^{(1)}$ computed in this way
agree with those computed from the results of Ref.~\cite{dfn}, where the $u$ 
quark is massless from the beginning.
The numerical results for the BLM corrections $A_{(0,j)}^{(2)}$ 
are very sensitive to the value of 
the charm quark mass employed in the code \cite{btocpert}, as at  
small $m_c$ the $A_{(0,j)}^{(2)}$ are proportional to $m_c^2 \ln^2 m_c$.
The numerical error associated with the 
choice of $m_c=50$~MeV in their computation is certainly acceptable 
for our purposes (it is  below 1\% in the BLM correction to the 
total rate, $A_{(0,0)}^{(2)}$, 
where it can be estimated using the exact result \cite{ritbergen}).
  In the case where only a  cut on the charged 
lepton energy is imposed,  it is also possible to express $M_{(i,j)}$ and 
$ A_{(i,j)}^{(1)}$ in compact analytic form; the expressions relevant
for the first three integer moments are given in  Appendix A and B. In general,
however, we rely on a numerical integration for the perturbative corrections.

As mentioned in the Introduction, we consider here truncated moments 
subject to a  lower cut on the energy of the charged lepton, $E_\ell^{\rm cut}$, 
and to an upper cut on the hadronic invariant mass, $M_X^{\rm cut}$. 
In the following we employ
\beq 
\xi = 2 \frac{E_\ell^{\rm cut}}{ m_b}.
\eeq

\begin{figure}[t]
\begin{center}
\mbox{\epsfig{file=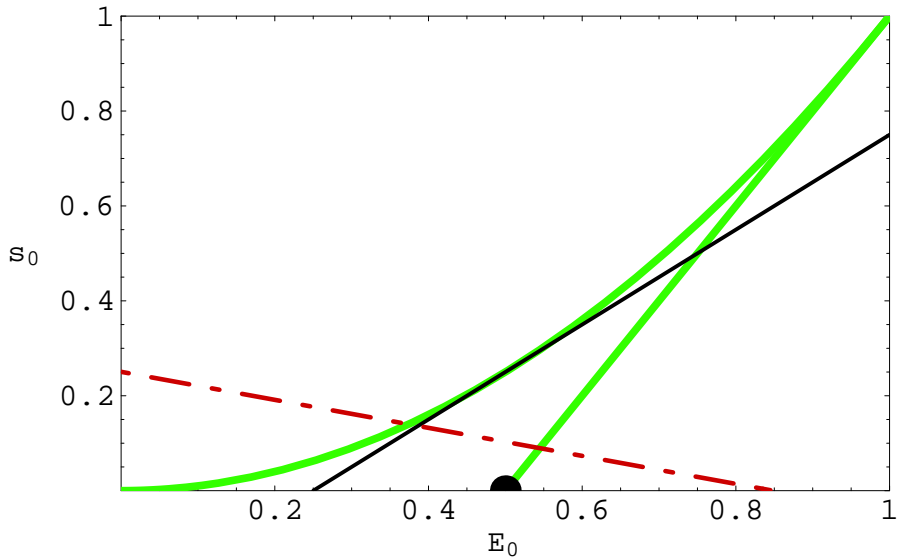,width=12cm}}
\end{center}
\caption{\sf 
Effect of lepton energy and $M_X$ cuts on the $E_0-s_0$ phase space. 
See the text for explanations. The cuts employed in the figure are 
 $M_X^{\rm cut} = 2.4$~GeV and $\xi = 0.5$.
} \label{cutplot}
\end{figure}

It is useful to explain the kinematics with cuts in some detail. 
The region of integration in the $E_0$-$s_0$ plane
is depicted in Fig.~\ref{cutplot}: the green (light) solid  
lines delimit the region of integration  without any cut, that is 
the region between the curves $s_0=2 E_0 -1$ and  $s_0 = E_0^2$.
The introduction of a cut in the lepton energy $E_\ell^{\rm cut}$
divides this region into three parts that should be treated 
differently (see {\it e.g.} \cite{falkluke}): in the figure 
these regions are separated by the black (dark)
 solid line that corresponds to
$s_0 = (1 - \xi) (2E_0 - 1 + \xi)$. 
In the first region, between $s_0 = 2 E_0 -1$ and 
$s_0 = (1 - \xi) (2E_0 - 1 + \xi)$ (below the black line), 
one should use the differential rate calculated with the electron energy 
cut imposed.  There are two regions above the black line, 
between $s_0 = (1 - \xi) (2E_0 - 1 + \xi)$ and $s_0 = E_0^2$.
In the lower of these two regions, 
the lepton energy is always above $E_\ell^{\rm cut}$,
and one should use the differential rate calculated without the cut.
Finally, the upper region above the black line is excluded as
the lepton energy is always  below the cut.
Whatever its value, a cut on the lepton energy affects both perturbative and
non-perturbative contributions to the moments.

A cut on the hadronic invariant mass $M_X^{\rm cut}$ 
limits the region of integration to the area below the red dash-dotted
line, that corresponds to   $M_X^2 = (M_X^{\rm cut})^2$.
Increasing the value of $M_X^{\rm cut}$, the allowed region of integration
expands.
For a  wide range of values of $M_X^{\rm cut}$ and $m_b$, 
as long as the red line does not come close to the $E_0$ axis to the left of 
$E_0=1/2$, the introduction of this cut affects only the perturbative 
corrections to the moments, as it excludes only events
characterized by high $s_0$ (hard gluon radiation).
 We limit ourselves to this case and
consider only  values of $M_X^{\rm cut}$ above the lower limit
\beq
\label{mxcut}
M_X^{\rm cut} > \sqrt{M_B {\bar \Lambda}}\, ,
\eeq
corresponding to the situation in which the cut in ${M_X^2}$ 
intersects the $E_0$ axis at the value $E_0=1/2$ (black dot).
Using $m_b = 4.6$~GeV, for instance, we have $M_X^{\rm cut} > 1.89$~GeV.
In fact, the effect of the $B$ distribution function becomes important within
distances of $O(\Lambda_{QCD})$ from the $E_0$ axis. As we will see in the 
next Section, a clean prediction of the moments requires $M_X^{\rm cut}$ 
significantly higher than in the above equation.
Table \ref{refpert} reports our results for the components of
${\cal M}_{(i,j)}$ in the on-shell mass scheme at particular $\xi$ and 
$M_X^{\rm cut}$ values.

\begin{table}
\begin{center}
\begin{tabular}{|c|c|l|l|l|l|}
\hline
i & j &~\hfill $M_{ij}$ \hfill~&~\hfill $A^{(1)}_{ij}$\hfill~&~\hfill 
$A^{(2)}_{ij}$ \hfill~\\
\hline
\hline
0  &  0  &   0.818792  &   -2.26120  &    -3.1225    \\
 \hline
 0  &  1  &   0.281010  &   -0.73861  &    -0.9206    \\
 \hline
 0  &  2  &   0.104107  &   -0.25651  &    -0.2921    \\
 \hline
 0  &  3  &   0.040583  &   -0.09238  &    -0.0944    \\
 \hline
 1  &  0  &  -0.004709  &    0.13096  &     0.2476    \\
 \hline
 1  &  1  &  -0.000878  &    0.05258  &     0.0979    \\
 \hline
 1  &  2  &   0.000058  &    0.02203  &     0.0405    \\
 \hline
 2  &  0  &   0.002061  &    0.00496  &     0.0076    \\
 \hline
 2  &  1  &   0.000878  &    0.00214  &     0.0032    \\
 \hline
 3  &  0  &   0.000115  &    0.00034  &     0.0005    \\
\hline
\end{tabular}
\end{center}
\caption{\sf \label{refpert} Various contributions to the building blocks 
in the on-shell scheme with $E_\ell>1$~GeV and $M_X<2.5$~GeV. 
The non-perturbative corrections in $M_{(i,j)}$ 
are calculated with the default values of the  parameters given in 
Eqs.~(\ref{default}) and  $X_{\mu} = 28$. }
\end{table}

Although we start from on-shell expressions \cite{btocpert,dfn}, 
we actually employ the Wilsonian scheme with a hard factorization scale
$\mu$, whose optimal value is close to 1 GeV \cite{kinetic}. 
 The Wilson coefficients in this scheme 
can be determined from the requirement that the observables 
be $\mu$-independent. The initial condition is that at $\mu\!\to\!0$ 
one should recover the results in the on-shell scheme.
In practice, at  low perturbative orders this often reduces
to re-expressing the pole-scheme results in terms of the running
$\mu$-dependent parameters.
In particular, the $\mu$-dependent parameters are the
 $b$ quark mass $m_b(\mu)$, kinetic expectation value $\mu_\pi^2(\mu)$, and
the Darwin expectation value $\rho_D^3(\mu)$\footnote{Unlike Ref.~\cite{btoc},
we employ here a running Darwin expectation value. The relation between $\tilde
\rho_D^3=\rho_D^3(0)$ and $\rho_D^3(\mu)$ can be found {\it e.g.} in 
\cite{imprecated}.}. In the following, our default choice for the 
non-perturbative parameters evaluated at $\mu=1$~GeV is
\bea \label{default}
&& \ \ \ \ \ \ m_b\!=\!4.6\GeV \ \ \ 
\mu_\pi^2\!=\!0.40\GeV^2 \ \ \
\mu_G^2\!=\!0.35\GeV^2 \\
&& \ \ \ \ \rho_D^3\!=\!0.1\GeV^3 \ \ \
\rho_{LS}^3\!=\!-0.1\GeV^3 \ \ \
\alpha_s(m_b)=0.22 \, .
\nonumber
\eea
The above values for the OPE parameters 
have been chosen having in mind the central values of the fits to 
$B\to X_c l \nu$  and $B\to X_s \gamma$ moments \cite{fits} 
and some additional constraint.

The coefficient function of the Darwin operator that contributes to 
 the $b\!\to\!q\,\ell\nu$ total width 
 at order $1/m_b^3$ has a logarithmic divergence \cite{1mb3,bds}
as $m_q\!\to\!0$:
\beq
C_D\simeq -\frac{\Gamma_0}{m_b^3} \left[8\ln\frac{m_b^2}{m_q^2}-\frac{77}{6}+
{\cal O}\left(\frac{m_q^2}{m_b^2}\right)\right] .
\label{darwin}
\eeq
The singularity originates from the domain of low-momentum 
final-state quark (i.e., large $q^2\!\simeq\! m_b^2$) and is  
removed by a one-loop penguin diagram that mixes the 
four-quark operator 
$O_{\scriptscriptstyle\rm WA}^u=  6 \bar b_L^\alpha \gamma_0 
b_L^\beta\,\bar{u}_L^\beta \gamma^0  u_L^\alpha
-2 \bar b_L^\alpha\vec\gamma 
b_L^\beta\,\bar{u}_L^\beta\vec\gamma  u_L^\alpha
$  
(here in its Fierzed form) into the Darwin operator.
Let us illustrate how this happens in the total semileptonic width:
the lowest order contribution of the Weak Annihilation (WA) operator
$O_{\scriptscriptstyle\rm WA}^u$ is \cite{WA,voloshin}
\beq
\delta \Gamma_{\scriptscriptstyle\rm WA}= \Gamma_0 \ C_{\scriptscriptstyle\rm WA} \ \langle B|O_{\scriptscriptstyle\rm WA}^u|B\rangle
\eeq
where $C_{\scriptscriptstyle\rm WA}=32\pi^2 / m_b^3$. In the factorization approximation
the matrix element
$B_{\scriptscriptstyle\rm WA}\equiv\langle B|O_{\scriptscriptstyle\rm WA}^u|B\rangle $ vanishes for $O_{\scriptscriptstyle\rm WA}$
corresponding to zero lepton masses. 
Including $O(\alpha_s)$ effects, the above equation  becomes
\bea
\delta \Gamma_{\scriptscriptstyle\rm WA}&=& \Gamma_0 \ C_{\scriptscriptstyle\rm WA} \left[
\left(1+O(\alpha_s)\right) B_{\scriptscriptstyle\rm WA}(\mu_{4q}) + 
\frac{\alpha_s}{\pi}\, a(\mu_{4q})
\langle B|\bar b \,t^a \,b \sum_q \bar q \,\gamma_0 \,t^a \,q|B\rangle
+...\right]\nonumber\\
&=& \Gamma_0  \left[C_{\scriptscriptstyle\rm WA}\,
 B_{\scriptscriptstyle\rm WA}(\mu_{4q}) - \frac{8\,\rho_D^3}{m_b^3} \, 
\ln \frac{m_u^2}{\mu_{4q}^2}+O(\alpha_s)\right],
\label{wa2}
\eea
where $\mu_{4q}$ is  the renormalization scale  of the WA operator 
and $a(\mu_{4q})$ is the contribution of  a penguin mixing diagram
  renormalized in the $\overline{\rm MS}$ scheme.
We have used the fact that the Darwin operator is 
proportional to  $g_s^2\,  \bar b  \,t^a\, b \sum_q \bar q 
\,\gamma_0\, t^a q$ by QCD equations of motion, 
and we have neglected those contributions proportional to the matrix
elements of  $O_{\scriptscriptstyle\rm WA}^u$ and of other operators 
that come with ${\cal O}(\alpha_s)$ Wilson coefficients, as they are 
irrelevant to the present discussion.
The constant accompanying the logarithm in Eq.~(\ref{wa2}) depends
on the renormalization scheme; it
vanishes in the $\overline{{\rm MS}}$ scheme that we have employed 
above\footnote{
This applies
if $O^u_{\scriptscriptstyle\rm WA}$ is expressed in its Fierzed form in the continuation
to  $D\ne 4$ dimensions, which is  part of the choice of scheme.
 Had we employed $O_{\scriptscriptstyle\rm WA}^u=-4 \bar b_L^\alpha\vec\gamma
u_L^\alpha\,\bar{u}_L^\beta\vec\gamma  b_L^\beta$  directly in $D$ dimensions,
using an anticommuting $\gamma_5$ (NDR scheme),
 the logarithm $\ln{m_b^2}/{\mu_{4q}^2}$
would be accompanied by a constant +2/3.}.
We have therefore seen that the inclusion of the WA operator
effectively replaces $\ln{m_b^2}/{m_u^2}$ in $C_D$ by $\ln{m_b^2}/{\mu_{4q}^2}$
plus a constant,
\beq
C_D =  -\frac{\Gamma_0}{m_b^3}
\left(8\ln\frac{m_b^2}{\mu_{4q}^2}-\frac{77}{6}\right)\,.
\label{darwin2}
\eeq
Varying the renormalization scale  $\mu_{4q}$ adds a piece
proportional to $\rho_D^3$ to the WA expectation value. 
This contribution is independent of the flavor of the spectator, 
though, and therefore does not affect the differences between 
$B^+$ and $B^0$.

In the following we assume factorization to hold at the scale 
$\mu_{4q}\!=\!0.8\GeV$, i.e.\ $B_{\scriptscriptstyle\rm 
WA}(0.8\,{\rm GeV})\!\simeq\!0$. A change in $\mu_{4q}$  sets the 
natural size of the non-factorizable contribution in 
$B_{\scriptscriptstyle\rm WA}$. To get a crude estimate of how 
the non-valence (flavor-singlet) 
non-factorizable component of  the expectation value of the WA operator
affects the OPE predictions,
we may vary  $\mu_{4q}$ in the interval  $0.4\GeV\!\lsim\!\mu_{4q}$
$\lsim\!1.7\GeV$. It is clear, however, that 
ultimately the size of the WA expectation values in 
both $B^0$ and $B^+$ must be determined experimentally.
Notice also that, unlike the total width,  the parton level moments
$M_{(i,j)}$ are not affected by the WA-Darwin
mixing for $i,j\neq 0$ (see Appendix A).

In the calculation of the moments we follow \cite{btoc} closely.
In particular, we consider $\lbar = {\cal O}(1)$ in 
the $\Lambda_{QCD}/m_b$ expansion and expand in $\Lambda_{QCD}/m_b$ and 
$\alpha_s$, neglecting all terms of $O(\alpha_s\ \Lambda_{QCD}^2/m_b^2)$.
This choice makes the hadronic moments sensitive to the choice of $\mu_{4q}$
in Eq.~(\ref{darwin2}). One can parameterize 
this dependence using 
$
X_{\mu} \equiv 8 \ln\frac{m_b^2}{\mu_{4q}^2},
$
and vary it  in the range $16 \leq X_{\mu}\leq 40$,
corresponding to the range in $\mu_{4q}$ just discussed, or equivalently 
use $|B_{\scriptscriptstyle\rm WA}(0.8~{\rm GeV})|\le 0.004\,{\rm GeV}^3$.
The contribution of the WA operator to the moments can be easily
recovered in the following 
by the replacement $\rho_D^3\, X_\mu \to\rho_D^3\, X_\mu -32\pi^2
 \,B_{\scriptscriptstyle\rm WA}(\mu_{4q})$.

In Table~\ref{eta2} we provide some reference numbers for $U_{1,2,3}$, 
obtained using $E_\ell^{\rm cut}=1$~GeV, $X_\mu = 28$, and the default values 
given in Eq.~(\ref{default}) at different values of $M_X^{\rm cut}$.
\begin{table}[t]
\centerline{
\begin{tabular}{|c|l|l|l|l|l|l|l|}\hline
$M_X^{\rm cut}$ & 2.3 & 2.5 &  2.7 & 3  & 3.5 & $M_B$ \\  \hline
$U_1$ & 1.898 & 1.960 & 1.997 & 2.028 & 2.045 &  2.047\\ \hline
$U_2$ & 1.724 & 1.997 &  2.062  &  2.228  &  2.357  & 2.377\\ \hline
$U_3$ &1.188 & 1.730  &  2.338 & 3.225 & 4.198 & 4.416\\ \hline
\end{tabular}} 
\caption{\sf \label{eta2}
Hadronic moments for  $ E_\ell^{\rm cut}= 1~\mbox{GeV}$, 
$B_{\scriptscriptstyle\rm WA}(0.8~{\rm GeV})=0$, and the default values given 
in Eq.~(\ref{default}) at different $M_X^{\rm cut}$. 
The results are in GeV to the appropriate power. }
\end{table}

\begin{table}[ht]
\centerline{  
\begin{tabular}{|c|l|l|l|l|l|l|l|l|}\hline 
$ E_\ell^{\rm cut}  $& ~\hfill $ V $\hfill~ & \hfill~ $B$ \hfill~ & ~\hfill
$ P $ \hfill~
& ~\hfill $G$ \hfill~&~\hfill $D$ \hfill~&~\hfill $L$ 
\hfill~&\hfill~ $S $ \hfill~  & ~\hfill $Y$ \hfill~\\ \hline
0& 2.179& -3.84& -0.75& 0.40& 
    0.84& -0.038& -2.76& 10.4\\ \hline
0.6& 
    2.143& -3.80& -0.76& 0.42& 
    0.85& -0.035& -2.83& 10.6\\ \hline
0.9& 
    2.078& -3.74& -0.80& 0.45& 
    0.89& -0.028& -2.96& 11.2\\ \hline
1.2& 
    1.973& -3.62& -0.89& 0.51& 
    0.96& -0.012& -3.16& 12.4\\ \hline
1.5& 
    1.818& -3.40& -1.04& 0.60& 1.10& 
    0.019& -3.45& 14.9\\ \hline
\end{tabular}}
\caption{\sf \label{tabU1} Coefficients of the linearized formulas in Eq.~(\ref{linearized})
for $U_1$ at different cuts on the lepton energy without $M_X$ cut.} 
\end{table}

\begin{table}[ht]
\centerline{ 
\begin{tabular}{|c|l|l|l|l|l|l|l|l|}\hline
$ E_\ell^{\rm cut}  $& ~\hfill $V$ \hfill~ & \hfill~ $B$ \hfill~ & ~\hfill
$ P $ \hfill~
& ~\hfill $G$ \hfill~&~\hfill $D$ \hfill~&~\hfill $L$ 
\hfill~&\hfill~ $S $ \hfill~  & ~\hfill $Y$ \hfill~\\ \hline
0 &  2.832 &  -0.706 &  5.11 &  -0.367 &  -5.01 &  -0.04 &  6.45 &  
    -20.8\\ 
\hline0.6 &  2.691 &  -0.681 &  
    4.98 &  -0.352 &  -5.07 &  -0.08 &  5.89 & -21.0\\ \hline
0.9 &  
    2.468 &  -0.452 &  4.75 &  -0.329 &  -5.22 &  -0.17 &  5.05 &  
    -21.5 \\ \hline
1.2 &  2.173 &  -0.368 &  
    4.39 &  -0.295 &  -5.51 &  -0.30 &  4.09 &  -22.9\\ \hline
1.5 &  
    1.835 &  -0.367 &  3.86 &  -0.247 &  -6.04 &  -0.47 &  3.33 &  
    -26.0\\ \hline
\end{tabular}}
\caption{\sf \label{tabU2}  Same as in Table~\ref{tabU1} but for $U_2 $.}
\end{table}

\begin{table}[ht] 
\centerline{
\begin{tabular}{|c|l|l|l|l|l|l|l|l|}\hline
$ E_\ell^{\rm cut}  $& ~\hfill $V$ \hfill~ & \hfill~ $B$ \hfill~ & ~\hfill
$P$ \hfill~
& ~\hfill $G$ \hfill~&~\hfill $D$ \hfill~&~\hfill $L$ 
\hfill~&\hfill~ $S$ \hfill~  & ~\hfill $Y$ \hfill~\\ \hline
0 &  7.096 &  1.156 &  4.72 &  0.150 &  20.9 &  1.3 &  
    21.0 &  35.8\\ \hline
0.6 &  6.119 &  1.131 &  4.62 &  0.140 &  20.1 &  
    1.2 &  15.6 &  35.7\\ \hline
0.9 &  4.872 &  0.689 &  4.55 &  0.113 & 
     18.6 &  1.1 &  8.66 &  35.6\\ \hline
1.2 &  3.519 &  0.239 &  4.52 & 
     0.067 &  16.2 &  1.0 &  1.39 &  35.9\\ \hline
1.5 &  
    2.342 &  -0.104 &  4.53 &  0.008 &  12.8 &  
    0.7 &  -4.30 &  37.3\\ \hline
\end{tabular}}
\caption{\sf \label{tabU3} Same as in Table~\ref{tabU1} but for $U_3 $.}
\end{table}

In general, the BLM corrections are almost as relevant and have the 
same sign as the one-loop 
perturbative contributions at fixed $\alpha_s=0.22$, {\it i.e.} they 
significantly decrease the effective scale of the QCD coupling.
It is also convenient to have approximate linearized formulas 
 for a generic moment ${\cal M}$ of the form
\bea\nonumber
{\cal M}(m_b,\mu_\pi^2,\mu_G^2, \rho_D^3,
\rho_{LS}^3;\alpha_s) \msp{-4}&=&\msp{-4}
V + B\,(m_b\!-\!4.6\GeV) + A\,(\alpha_s-0.22)  \;\\
&&\msp{-.9}+ P\,(\mu_\pi^2\!-\!0.4\GeV^2)
+D\,(\rho_D^3\!-\!0.1\GeV^3) \qquad 
\label{linearized}\\
\nonumber
&&\msp{-.9}+  G\,(\mu_G^2\!-\!0.35\GeV^2)
+L\,(\rho_{LS}^3\!+\!0.1\GeV^3)\\ \nonumber
&&\msp{-.9}+Y\, B_{\scriptscriptstyle\rm WA}(0.8~{\rm GeV}); 
\eea
The values of $V$ are obtained with the  default values of the heavy quark 
parameters, $B_{\scriptscriptstyle\rm WA}(0.8~{\rm GeV})=0$,
and are  quoted in $\GeV$ to the corresponding power.
In Tables~\ref{tabU1}-\ref{tabU3}   we report values of the various 
coefficients with  different $E_\ell^{\rm cut}$ in the case without an $M_X$ cut.
For values of $M_X^{\rm cut}$ satisfying the bound in Eq.~(\ref{mxcut})
only the perturbative contributions differ from the case without an $M_X$ cut. 
Table~\ref{tabmxcut2} therefore shows only the coefficients 
$V$, $B$, and $S$ for $M_X^{\rm cut}=2.5$~GeV and $E_\ell^{\rm cut}=0.9$~GeV.
The results of our calculation are implemented in  a FORTRAN code, available 
from the authors, that computes hadronic moments in $b\to u$ for arbitrary 
$E_\ell^{\rm cut}$ and for $M_X^{\rm cut}$ satisfying Eq.~(\ref{mxcut}).

\begin{table}[ht]
\centerline{
\begin{tabular}{|c|l|l|l|}\hline
&~\hfill $V$ \hfill~&~\hfill $B$ \hfill~&~\hfill  $S$ \hfill~\\  \hline
$U_1$ &1.981& -3.56& -3.6 \\ \hline
$U_2$& 1.939& 0.10& 1.6\\ \hline
$U_3$& 1.743& 1.60& -11.7\\ \hline
\end{tabular}}
\caption{\sf  Coefficients of the linearized formulas in Eq.~(\ref{linearized})
for $U_{1,2,3}$ for $M_X^{\rm cut}=2.5$~GeV and $E_\ell^{\rm cut}=0.9$~GeV. 
Only the coefficients that differ from Tables~\ref{tabU1}-\ref{tabU3} are 
reported.}
\label{tabmxcut2}
\end{table}

We can compare the $U_1$ values given in Tables~\ref{tabU1} and \ref{tabmxcut2}
to the preliminary BaBar results of Ref.~\cite{babarspectrum}.
With a high $M_X$ cut of 5~GeV BaBar find $U_1(5~{\rm GeV})=2.78\pm0.82$~GeV$^2$, 
that is in good agreement with our reference value of 2.18~GeV$^2$.
BaBar also reports a result at low $M_X^{\rm cut}=1.86$~GeV which is 
very close to the lower bound of Eq.~(\ref{mxcut}). In that case
their result $U_1(1.86~{\rm GeV})=1.98\pm0.20$~GeV$^2$ 
is compatible with the reference value 1.49~GeV$^2$ 
that we obtain at $M_X^{\rm cut}=1.9$~GeV, although Fermi motion effects, that 
shift $U_1$ to higher values, 
have not been included in the calculation (see next Section).

Finally, using our building blocks 
it is straightforward to study also $q^2$ moments. Indeed, 
replacing $\bar\Lambda$ with $-m_b$ in the rhs of Eq.~(\ref{mx2}) 
one obtains $q^2$ and can then calculate the $q^2$ moments 
\beq
\langle q^{2n}\rangle = \frac{\int d  q^2 \ q^{2n}\  d\Gamma/d q^2}
{\int d  q^2\  d\Gamma/d q^2}
\eeq
in a way similar to the invariant mass moments.
Table~\ref{tabq2} gives the reference values and the coefficients of 
the linearized formula of Eq.~(\ref{linearized})
 for the first three
moments in the case of $E_\ell^{\rm cut}=1.2$~GeV and no cut on $M_X$.
\begin{table}[t]
\centerline{
\begin{tabular}{|c|r|r|r|r|r|r|r|r|}\hline
&~\hfill $V$ \hfill~&~\hfill $B$\hfill~&~\hfill $P$ \hfill~&~\hfill $G$ 
\hfill~&~\hfill $D$ \hfill~&~\hfill $L$ 
\hfill~&~\hfill $S $ \hfill~&~\hfill $10^3\,Y$ \hfill~\\\hline
 $\langle q^2\rangle$ &
7.773 &  3.058 &  -0.0193 &  -0.894 &  -0.710 &  -0.186 &  6.15 &  -0.084\\ \hline
 $\langle q^4\rangle$ & 
81.32 &  71.35 &  -0.188 &  -19.7 &  -4.26 &  -4.22 &  113.6 & -2.27\\ \hline
 $\langle q^6\rangle$ &
980.4 &  1461 &  -1.417 &  -381 &  201 &  -82.9 &  1758 &  -52.2\\ \hline
\end{tabular}
}
\caption{\sf  Coefficients of the linearized formulas in Eq.~(\ref{linearized})
for $\langle q^{2,4,6}\rangle$ with  $E_\ell^{\rm cut}=1.2$~GeV. \label{tabq2}}
\end{table}

\section{Theoretical uncertainty}
Let us now consider the various sources of theoretical uncertainty that
affect our predictions. First we consider the uncertainty that affects 
the moments when no upper cut on $M_X$ is imposed. 
If the $E_\ell$ cut is not too severe 
(less than, say, 1.4~GeV), there are four main theoretical systematics:
\begin{itemize}
\item
[i)] uncalculated  $O(\alpha_s^2)$ and $O(\alpha_s \Lambda^{2,3}_{QCD}/m_b^{2,3})$
perturbative contributions to the Wilson coefficients; 
\item
[ii)] missing $O(1/m_b^4)$ non-perturbative effects;
\item 
[iii)] the error from the scale in  $X_\mu$;
\item
[iv)] Weak annihilation (WA) contributions.
\end{itemize}
The first two items are common with the $b\to c$ moments and can be analyzed 
in a similar way. 
The last two items, as clarified in the previous Section, are
two facets of the same effect, and should not be counted twice. When
we include the WA effects as a priori unknown, we use the variation of
the scale $\mu_{4q}$ in iii) to estimate the size of its
flavor singlet contributions (flavor non-singlet WA effects 
can be studied from the difference in the moments for charged and
neutral $B$). On the other hand, as discussed below, 
moment measurements allow to place
constraints or to detect WA. In this approach fixing $\mu_{4q}$ in
$X_\mu$ extracts the expectation value of the WA operator normalized
(in dimensional regularization) at this point. The remaining
uncertainty comes from higher-order corrections to the Wilson
coefficients, item i).

In general, we note that $b\to u$ moments are
affected by larger theory errors than the $b\to c$ moments.
Using the simple recipe given in \cite{btoc}, we estimate the 
uncertainty related to i) and ii) above 
by varying $\mu_\pi^2$, 
$\mu_G^2$, and $\alpha_s$ by $\pm20$\%, $\rho_D^3$ 
and $\rho_{LS}^3$ by $\pm$30\%, and $m_b$ by $\pm$20~MeV 
in the theoretical predictions in an uncorrelated way.
The  typical results 
\beq
\label{errors1}
\delta U_1/U_1\approx 8\% \ \ \ 
\delta U_2/U_2\approx 20\% \ \ \ 
\delta U_3/U_3 \approx 20\%, 
\eeq
roughly reflect  the theory errors due to i)  and ii) above, independently
of the accuracy with which we know the OPE parameters from \cite{fits}.
They are driven by the strong sensitivity of $U_2$ and $U_3$ 
to $\mu_\pi^2$ and $\rho_D^3$. 

The uncertainty of Eq.~(\ref{errors1}) 
can be estimated in alternative ways.
For instance, we can evaluate $U_{1,2,3}$ using a different rearrangement of 
non-perturbative corrections, namely considering 
$\bar\Lambda$ as an $O(\Lambda_{QCD})$ quantity in the expansion in inverse 
powers of the $b$ mass. In this case the moments are insensitive to
$X_\mu$ and to the related error. The results are always within the ranges 
in Eq.~(\ref{errors1}).
The main step necessary  to improve on 
the  above uncertainties is the calculation of perturbative contributions to 
the Wilson coefficients, of $O(\alpha_s^2)$ \ \footnote{They are  already 
available for the $q^2$ spectrum and moments \cite{q2spectrum}, 
as well as for the total rate (first paper of \cite{ritbergen}).} 
 and $O(\alpha_s/m_b^{2,3})$.

For what concerns the value of $B_{\scriptscriptstyle\rm WA}$, 
as mentioned in the previous 
section we vary it in the range $|B_{\scriptscriptstyle\rm WA}(0.8~{\rm GeV})|\le 0.004~{\rm GeV}^3$.
This is a rather conservative estimate for the a priori unknown
flavor singlet WA contribution, that induces a typical uncertainty of 
about 2\%, 3\%, 2\% for $U_1$, $U_2$, $U_3$, respectively, although the error
can be larger for high $E_\ell$ cuts, as it is evident from 
Tables~\ref{tabU1}-\ref{tabU3}. 

We recall that WA contributions  are concentrated at maximal $q^2$, namely at 
the origin in Fig.~1. In the $(E_0,s_0)$ plane the fixed $q^2$
contours are straight lines identified by $s_0= 2E_0-1+q^2/m_b^2$. 
For instance, the r.h.s.\ boundary of the 
relevant phase space (the straight green line) corresponds to $q^2=0$.
WA contributions are therefore characterized by small hadronic invariant
mass and can be relevant in the total rate, but are 
suppressed in the moments of $M_X^2$.

Turning  to the theoretical uncertainty  introduced by a cut on $M_X$,
we have already mentioned in the previous section that 
low $M_X^{\rm cut}$ make the 
moments sensitive to the $B$ distribution function.
A rough but simple way to understand the range of $M_X^{\rm cut}$ for which these 
non-perturbative effects become important is to rewrite 
Eq.~(\ref{mxcut}) shifting  $\bar\Lambda$ by 
the typical width of the distribution function, {\it i.e.}\ 
$\pm \sqrt{ \mu^2_\pi/3}\approx \pm 0.37$~GeV, to account for the 
Fermi motion. The result is that 
distribution function effects become important when 
$M_X^{\rm cut}$ is less  than about 2.35~GeV.
This result, however,  is unlikely to apply to higher moments that are more 
sensitive to the tail of the distribution function.
A detailed estimate would imply a dedicated implementation of the 
distribution function, which is beyond the scope of the present publication.

A detailed study of similar effects in the photon energy
moments of $B\!\to\!X_s+\gamma$  \cite{Benson:2004sg}
showed that if the Wilsonian-type OPE
with the hard cutoff at the scale around $1\GeV$ is used, 
the inclusion of perturbative corrections affects
only marginally the bias induced by Fermi motion.
At the same time, the estimates are sensitive to
the  $1/m_b$ corrections that decrease  the variance of
the  light-cone distribution with respect to the heavy quark limit.
Based on that experience, we have estimated the Fermi motion effects
introduced by  the $M_X$ cut in the $b\!\to\!u \, \ell\nu$
moments by smearing the tree-level
differential rate with an exponential distribution function
(see eq.(13) of Ref.~\cite{Benson:2004sg})
characterized by the low value $\mu_\pi^2\!\approx\!
0.36\GeV^2$,  to approximately account for the $1/m_b$ contributions
to the second moment of the distribution function.
Since the calculation of the hadronic moments is  sensitive
to the tail of the distribution, this estimate depends critically
on the  functional form adopted. In this respect, our choice
of the exponential form leads to more conservative estimates
than, say, with  a Gaussian ansatz.
Comparing the moments at $M_X^{\rm cut}=M_B$ to  those at various 
$M_X^{\rm cut}$,  we find that the Fermi motion may alter $U_1$ by 
$\sim 6\%$ at $M_X^{\rm cut}=2.5$~GeV and by  $\sim 3\%$ at $M_X^{\rm cut}=2.7$~GeV.
The higher moments are of course more sensitive: $U_2$ may vary 
by $30\%$ at $M_X^{\rm cut}=2.5$~GeV and by  $\sim 20\%$ at $M_X^{\rm cut}=2.7$~GeV, 
while $U_3$ is dramatically affected -- $O(1)$ effects -- 
below $M_X^{\rm cut}=$2.7~GeV; the Fermi motion effects might become 
comparable to the other uncertainties only above $M_X^{\rm cut}=3$~GeV.
Such high $M_X^{\rm cut}$ values are certainly challenging for 
present experiments,
but preliminary studies indicate the possibility of measuring the moments with
interesting accuracy even for $M_X^{\rm cut}$ well above the charm threshold 
\cite{tackmann}.

Finally, let us discuss the uncertainty in the evaluation of the $q^2$ moments.
We have already mentioned that these moments are very sensitive to
the WA contributions, and therefore can be used to detect or place
constraints on them. The flavor non-singlet WA contributions 
can be studied  by comparing the $q^2$ moments (or other decay characteristics)
measured in charged and neutral $B$ decays separately \cite{WA}.
For precision studies it is important to control the flavor singlet
component of WA as well. They can be constrained by comparing 
the OPE predictions with data. In that respect, we need to consider only
the uncertainties listed under i) and ii); regarding them in 
the same way as for the hadronic moments, the results are
\beq 
\delta \langle q^2\rangle/\langle q^2\rangle\approx 3\% \ \ \
\delta \langle q^4\rangle/\langle q^4\rangle\approx 6\% \ \ \
\delta \langle q^6\rangle/\langle q^6\rangle\approx 9\% 
\eeq
and are dominated by perturbative effects.
As mentioned in the previous section, 
the sensitivity of the various moments to the WA contributions can be
understood from their $X_\mu$ dependence: varying 
$B_{\scriptscriptstyle\rm WA}(0.8~{\rm GeV})$ in 
the usual range and using the linearized formula of Eq.~(\ref{linearized}), 
we obtain a shift of approximately $4\%$, 11\%, 21\%, for the first three 
$q^2$ moments, respectively. 
Since we only consider $q^2$ moments without cuts on $M_X$, we do not have 
Fermi motion effects.

\section{Summary}
We have calculated the first three moments of the hadronic invariant mass
distribution in charmless semileptonic decays in a Wilsonian scheme 
characterized by a hard cutoff $\mu\sim 1$~GeV. Our calculation includes all
known perturbative and non-perturbative effects, through $O(\alpha_s^2\beta_0)$
and $O(1/m_b^3)$ and is implemented in a FORTRAN code available from the 
authors. As required by the present experimental situation,
we have considered  cuts on the lepton energy and on the 
invariant hadronic mass and have obtained approximate formulas that 
summarize the dependence of the moments on the OPE parameters.

The theoretical uncertainty of our OPE predictions ranges from 5\% to
30\%, but an  upper cut on $M_X$ introduces a dependence on the Fermi motion
of the $b$ quark in the meson. While we have performed a first 
estimate of these effects, the subject requires  a more detailed 
investigation that we postpone to a future publication. Moreover, 
as the constraints on the shape of the distribution function are likely 
to improve in the future, our estimates of the  Fermi motion effects
should not be considered as an irreducible uncertainty. Conversely,
the $M_X$ spectrum and its truncated moments can themselves be used to 
constrain the distribution function, especially in the tail that 
is not accessible in radiative decays. 

We find that the bias introduced by the distribution function
is not important for the first 
hadronic moment, if  the $M_X$ cut is placed above 2.5--2.7~GeV.
In that case its theoretical uncertainty is in the  10\% range. 
The prediction for the second central moment, $U_2$,
is subject to a 20\% uncertainty even without a cut on $M_X$.
For cuts on $M_X$ higher than $\sim 2.7~$GeV, the Fermi motion
uncertainty on $U_2$ may be as high as 20\%.
Finally, the third central moment,  $U_3$, is very sensitive to Fermi motion
and can be predicted with a meaningful accuracy ($\sim 20\%$) only if
the cut on $M_X$ is higher than $\sim 3.2~$GeV.

 We have also considered $q^2$ moments which could soon be measurable 
and are particularly sensitive to WA.  They can be predicted with good 
accuracy in the local OPE, provided the cut on the lepton energy is 
sufficiently low, $E_\ell^{\rm cut}\lsim 1.5$~GeV. We have shown how it is 
possible to constrain
both the flavor  singlet and the flavor non-singlet WA contributions.
In view of their potential interest, 
they should therefore be measured with $E_\ell^{\rm cut}$ as low  as possible.

\section*{Acknowledgments}
We are grateful to Marco Battaglia, Oliver Buchmuller,  
Riccardo Faccini, Paolo Giordano, Bob Kowalewski,
Giovanni Ridolfi, and Kerstin Tackmann for helpful
discussions.  The work of P.~G.\ and G.~O.\ 
is supported in part by the EU grant
MERG-CT-2004-511156 and by MIUR under contract 2004021808-009, and
that of N.~U.\ by the NSF under grant PHY-0087419.

\section*{Appendix A: non-perturbative corrections}
Here we give explicit expressions for the lowest order and OPE contributions
to $M_{ij}$, in the case of a lower cut on the charged lepton energy.
They agree with Ref.~\cite{fls} for $\xi=0$.

\bea
M_{(0,0)} & = & 1+\xi ^4-2 \xi ^3  
-\frac{1}{6} \frac{\mu^2_{G}}{m_b^2} \left(5 \xi ^4+8
   \xi ^3+9\right)
+\frac{1}{6} \frac{\mu^2_{\pi}}{m_b^2} \left(5 \xi ^4-3\right)
 +\frac{\rho^3_{LS}}{m_b^3} \frac{\xi^4+9}{6}\nonumber \\ 
  & & +\frac{\rho^3_{D}}{m_b^3}
   \left(\frac{\xi ^4}{6}+\frac{8 \xi ^3}{3}-4 \xi ^2-8
   \xi - 8 \ln (1-\xi
   )  +\frac{77}{6}-X_\mu\right) 
\nonumber
\eea

\bea
M_{(0,1)} & = & \frac{-2 \xi ^5+15 \xi ^4-20 \xi
   ^3+7}{20} +\frac{1}{12} \frac{\mu^2_{G}}{m_b^2} \left(\xi^2-
4 \xi -6\right) \xi^3
-\frac{1}{12} \frac{\mu^2_{\pi}}{m_b^2} \left(\xi^5-6 \xi ^4-2
   \xi ^3+6\right)  \nonumber \\ 
  & & + \frac{1}{60} \frac{\rho^3_{LS}}{m_b^3} \left(-\xi ^5+10 \xi
   ^4+30 \xi ^3+6\right)+\frac{1}{60} \frac{\rho^3_{D}}{m_b^3} \left(-\xi ^5+10
   \xi ^4+90 \xi ^3-134\right)\nonumber
\eea

\bea
M_{(0,2)} & = & \frac{1}{60} \left(\xi ^6-9 \xi ^5+30 \xi
   ^4-30 \xi ^3+8\right)+\frac{1}{360} \frac{\mu^2_{G}}{m_b^2} \left(-5 \xi ^6+24 \xi ^5-60 \xi
   ^4-60 \xi ^3+26\right)\nonumber \\ 
  & & +\frac{1}{360} \frac{\mu^2_{\pi}}{m_b^2} \left(5 \xi^6-36 \xi ^5+120 \xi ^4-86\right)\nonumber
 +\frac{1}{360} \frac{\rho^3_{D}}{m_b^3} \left(\xi ^6-12
\xi^5+30 \xi ^4+120 \xi ^3-48\right)\\ 
& & + \frac{1}{360} \frac{\rho^3_{LS}}{m_b^3}
   \left(\xi ^6-12 \xi ^5+30 \xi ^4+120 \xi^3-28\right)\nonumber 
\eea

\bea
M_{(0,3)} & = & \frac{-2 \xi ^7+21 \xi ^6-84 \xi ^5+175 \xi ^4-140 \xi
   ^3+30}{560}\nonumber \\ 
  & & + \frac{\mu^2_{G}}{m_b^2} \frac{ \left(5 \xi ^7-28 \xi
   ^6+84 \xi ^5-175 \xi ^4-70 \xi
   ^3+72\right)}{1680}\nonumber\\
  & &
 + \frac{\mu^2_{\pi}}{m_b^2} \frac{ \left(-5 \xi ^7+42 \xi ^6-168 \xi ^5+490
   \xi ^4-210 \xi ^3-156\right)}{1680}\nonumber \\ 
  & & +\frac{\rho^3_{LS}}{m_b^3} \frac{ \left(-\xi ^7+14 \xi ^6-42 \xi
   ^5+140 \xi ^4+210 \xi ^3-104\right)}{1680}\nonumber \\ 
  & & +  \frac{\rho^3_{D}}{m_b^3} \frac{
   \left(-\xi ^7+14 \xi ^6-42 \xi ^5+420 \xi ^4-70 \xi
   ^3-48\right)}{1680}\nonumber  
\eea

\bea
M_{(1,0)} & = & \frac{1}{12} \frac{\mu^2_{G}}{m_b^2} \left(-2 \xi ^5-5 \xi ^4+4
   \xi ^3+3\right)+\frac{1}{60} \frac{\mu^2_{\pi}}{m_b^2} \left(14 \xi
   ^5-15 \xi ^4+20 \xi ^3-39\right)\nonumber \\  
  & &
+\frac{1}{60} \frac{\rho^3_{LS}}{m_b^3} \left(-2 \xi ^5-35 \xi ^4+60 \xi
   ^3-3\right)
+ \frac{1}{60} \frac{\rho^3_{D}}{m_b^3} \left(-2 \xi ^5+65 \xi
   ^4-140 \xi ^3+177\right)\nonumber
\eea

\bea
M_{(2,0)} & = &\frac{1}{45} \frac{\mu^2_{\pi}}{m_b^2}
   \left(\xi ^6-9 \xi ^5+30 \xi ^4-30 \xi
   ^3+8\right)
+ \frac{1}{45} \frac{\rho^3_{D}}{m_b^3} \left(-2 \xi ^6+9 \xi ^5-15 \xi
   ^4+30 \xi ^3-31\right)\nonumber \\ 
  & &+\frac{1}{45} \frac{\rho^3_{LS}}{m_b^3} \left(-\xi ^6-15
   \xi ^5+45 \xi ^4-30 \xi ^3+1\right)\nonumber
\eea

\bea
M_{(3,0)} & = & \frac{1}{210} \frac{\rho^3_{D}}{m_b^3} \left(-2 \xi ^7+21 \xi ^6-84 \xi
   ^5+175 \xi ^4-140 \xi ^3+30\right)\nonumber
\eea

\bea
M_{(1,1)} & = & \frac{1}{180} \frac{\mu^2_{G}}{m_b^2} \left(5 \xi ^6-3 \xi
   ^5-45 \xi ^4+30 \xi ^3+13\right)+
\frac{1}{180} \frac{\mu^2_{\pi}}{m_b^2} \left(-7 \xi ^6+27 \xi ^5+15 \xi
   ^4-30 \xi ^3-23\right)\nonumber \\ 
  & &+\frac{1}{180} \frac{\rho^3_{D}}{m_b^3} \left(\xi ^6-21
   \xi ^5+165 \xi ^4-150 \xi ^3+71\right)
+\frac{1}{180} \frac{\rho^3_{LS}}{m_b^3}\left(\xi ^6+15 \xi ^5-15 \xi ^4+30 \xi
   ^3-13\right)\nonumber
\eea

\bea
M_{(2,1)} & = & \frac{1}{420} \frac{\mu^2_{\pi}}{m_b^2} \left(-2 \xi ^7+21 \xi ^6-84 \xi
   ^5+175 \xi ^4-140 \xi ^3+30\right)\nonumber \\ 
  & &+\frac{1}{420} \frac{\rho^3_{D}}{m_b^3}
   \left(4 \xi ^7-21 \xi ^6+28 \xi ^5+35 \xi
   ^4-74\right)\nonumber \\ & &
+  \frac{\rho^3_{LS}}{m_b^3}  \frac{ \left(6 \xi ^7+77 \xi ^6-420 \xi
   ^5+735 \xi ^4-420 \xi ^3+22\right)}{1260}\nonumber
\eea

\bea
M_{(1,2)} & = &\frac{\mu^2_{G}}{m_b^2} \frac{ \left(-10 \xi ^7+21 \xi ^6+56 \xi ^5-245
   \xi ^4+140 \xi ^3+38\right)}{1680}\nonumber \\ 
  & &+\frac{1}{240} \frac{\mu^2_{\pi}}{m_b^2} \left(2 \xi^7-11
   \xi ^6+8 \xi ^5+55 \xi ^4-60 \xi ^3-2\right)\nonumber\\
  & &+\frac{\rho^3_{LS}}{m_b^3} \frac{ \left(-6
   \xi ^7-91 \xi ^6+84 \xi ^5+735 \xi ^4-420 \xi
   ^3-134\right)}{5040}\nonumber \\ 
  & &+\frac{\rho^3_{D}}{m_b^3} \frac{ \left(-2 \xi ^7+35 \xi
   ^6-364 \xi ^5+1155 \xi ^4-700 \xi
   ^3+30\right)}{1680}\nonumber 
\eea

\section*{Appendix B: ${\cal O}(\as)$ corrections to the moments}

We report here analytic formulas for the ${\cal O}(\as)$ perturbative 
corrections to the building blocks $M_{ij}$ defined in Eq.~(\ref{Mij})
when a lower cut on the lepton energy is applied. The expressions are valid in 
the on-shell scheme for the $b$ quark mass and they agree
with Ref.~\cite{fls} for $\xi=0$.
We employ the short-hand $L_\xi= \ln (1-\xi )$.

\bea
A^{(1)}_{(0,0)} & = & -\frac{4}{9} \pi ^2 \xi ^4+\frac{47 \xi ^4}{18}+\frac{8
   \pi ^2 \xi ^3}{9}-\frac{59 \xi ^3}{9}+\frac{50 \xi
   ^2}{9}-\frac{52 \xi }{9}-\frac{2 \pi ^2}{3}+\frac{25}{6}
  + \frac{4 \xi^3}{3} \left(2 -\xi\right)\mbox{Li}_2(\xi)\nonumber  \\ 
& & +L_\xi \left(-\frac{8 \xi
   ^4}{9}+\frac{32 \xi ^3}{9}-\frac{14 \xi
   ^2}{3}+\frac{70 \xi
   }{9}-\frac{52}{9}\right)
+L_\xi^2 \left(-\frac{2
   \xi ^4}{3}+\frac{4 \xi
   ^3}{3}-\frac{2}{3}\right)  \nonumber 
\eea

\bea
A^{(1)}_{(0,1)} & = & \frac{2 \pi ^2 \xi ^5}{45}-\frac{103 \xi
   ^5}{450}-\frac{\pi ^2 \xi ^4}{3}+\frac{7 \xi
   ^4}{4}+\frac{4 \pi ^2 \xi ^3}{9}-\frac{1213 \xi
   ^3}{360}+\frac{191 \xi ^2}{72}-\frac{421 \xi
   }{180} \nonumber \\ \nonumber 
& & +L_\xi \left(\frac{4 \xi
   ^5}{45}-\frac{91 \xi ^4}{180}+\frac{91 \xi
   ^3}{90}-\frac{83 \xi ^2}{45}+\frac{323 \xi
   }{90}-\frac{421}{180}\right)\\ \nonumber
& & +L_\xi^2 \left(\frac{\xi ^5}{15}-\frac{\xi
   ^4}{2}+\frac{2 \xi
   ^3}{3}-\frac{7}{30}\right) 
+ \mbox{Li}_2(\xi ) \left(\frac{2 \xi
   ^5}{15}-\xi ^4+\frac{4 \xi ^3}{3}\right)
   -\frac{7 \pi ^2}{30}+\frac{1381}{900}
\eea

\bea
A^{(1)}_{(0,2)} & = & -\frac{1}{135} \pi ^2 \xi ^6+\frac{329 \xi
   ^6}{8100}+\frac{\pi ^2 \xi ^5}{15}-\frac{497 \xi
   ^5}{1350}-\frac{2 \pi ^2 \xi ^4}{9} \nonumber \\ \nonumber
& & +\frac{3757 \xi
   ^4}{2700}+\frac{2 \pi ^2 \xi ^3}{9}-\frac{71057 \xi
   ^3}{32400}+\frac{1121 \xi ^2}{675}-\frac{782 \xi
   }{675}-\frac{4 \pi^2}{45}+\frac{2257}{3600}\\ \nonumber 
& & +L_\xi \left(-\frac{2 \xi ^6}{135}+\frac{19
   \xi ^5}{180}-\frac{37 \xi ^4}{180}+\frac{73 \xi
   ^3}{180}-\frac{77 \xi ^2}{60}+\frac{484 \xi
   }{225}-\frac{782}{675}\right)\\ \nonumber 
& & +L_\xi^2
   \left(-\frac{\xi ^6}{90}+\frac{\xi ^5}{10}-\frac{\xi
   ^4}{3}+\frac{\xi
   ^3}{3}-\frac{4}{45}\right) \nonumber 
+ \mbox{Li}_2(\xi )   \left(-\frac{\xi
   ^6}{45}+\frac{\xi ^5}{5}-\frac{2 \xi ^4}{3}+\frac{2
   \xi ^3}{3}\right) 
\eea

\bea
A^{(1)}_{(0,3)} & = & \frac{\pi ^2 \xi ^7}{630}-\frac{5231 \xi
   ^7}{529200}-\frac{\pi ^2 \xi ^6}{60}+\frac{5237 \xi
   ^6}{50400}+\frac{\pi ^2 \xi ^5}{15}-\frac{24163 \xi
   ^5}{50400}-\frac{5 \pi ^2 \xi ^4}{36}+\frac{123257 \xi^4}{100800} \nonumber 
\\ \nonumber
& & +\frac{\pi ^2 \xi ^3}{9}-\frac{49307 \xi
   ^3}{30240}+\frac{116689 \xi ^2}{100800}-\frac{32143
   \xi }{50400}
   -\frac{\pi ^2}{28}+\frac{289223}{1058400}
\\ \nonumber
& & +L_\xi \left(\frac{\xi
   ^7}{315}-\frac{137 \xi ^6}{5040}+\frac{99 \xi
   ^5}{1400}-\frac{361 \xi ^4}{3360}+\frac{205 \xi
   ^3}{504}-\frac{2413 \xi ^2}{2100}+\frac{9077 \xi
   }{6300}-\frac{32143}{50400}\right)\\ \nonumber
& & +L_\xi^2 \left(\frac{\xi
   ^7}{420}-\frac{\xi ^6}{40}+\frac{\xi ^5}{10}-\frac{5
   \xi ^4}{24}+\frac{\xi
   ^3}{6}-\frac{1}{28}\right) \nonumber 
+  \mbox{Li}_2(\xi)    \left(\frac{\xi
   ^7}{210}-\frac{\xi ^6}{20}+\frac{\xi ^5}{5}-\frac{5
   \xi ^4}{12}+\frac{\xi ^3}{3}\right) 
\eea

\bea
A^{(1)}_{(1,0)} & = & \frac{\xi ^5}{50}-\frac{7 \xi ^4}{45}-\frac{37 \xi
   ^3}{90}+\frac{127 \xi ^2}{90}-\frac{16 \xi
   }{15} +\frac{91}{450}  \nonumber 
+L_\xi \left(-\frac{\xi ^5}{10}+\frac{8 \xi
   ^4}{9}-\frac{5 \xi ^3}{3}+\frac{35 \xi
   }{18}-\frac{16}{15}\right)
\eea

\bea
A^{(1)}_{(2,0)} & = & \frac{31 \xi ^6}{4050}-\frac{11 \xi ^5}{135}+\frac{1207
   \xi ^4}{2700}-\frac{1912 \xi ^3}{2025}+\frac{574 \xi
   ^2}{675}-\frac{199 \xi }{675}+\frac{5}{324}\nonumber\\ \nonumber
& & +L_\xi \left(\frac{13
   \xi ^6}{1350}-\frac{13 \xi ^5}{150}+\frac{4 \xi
   ^4}{45}+\frac{61 \xi ^3}{135}-\frac{7 \xi
   ^2}{6}+\frac{449 \xi
   }{450}-\frac{199}{675}\right)
\eea

\bea
A^{(1)}_{(1,1)} & = & \frac{7 \xi ^6}{900}-\frac{31 \xi ^5}{450}+\frac{301 \xi
   ^4}{900}-\frac{202 \xi ^3}{225}+\frac{1007 \xi
   ^2}{900}-\frac{263 \xi }{450}+\frac{9}{100}\nonumber\\ \nonumber
& & +L_\xi
   \left(\frac{\xi ^6}{60}-\frac{77 \xi ^5}{450}+\frac{5
   \xi ^4}{9}-\frac{19 \xi ^3}{45}-\frac{29 \xi
   ^2}{36}+\frac{127 \xi
   }{90}-\frac{263}{450}\right)
\eea

\bea
A^{(1)}_{(2,1)} & = & -\frac{937 \xi ^7}{264600}+\frac{2791 \xi
   ^6}{75600}-\frac{103 \xi ^5}{504}+\frac{22087 \xi
   ^4}{37800}-\frac{32167 \xi ^3}{37800}+\frac{1027 \xi
   ^2}{1680}-\frac{2291 \xi }{12600}+\frac{1081}{132300}\nonumber\\ 
& & +L_\xi
   \left(-\frac{13 \xi ^7}{6300}+\frac{41 \xi
   ^6}{1800}-\frac{23 \xi ^5}{450}-\frac{43 \xi
   ^4}{360}+\frac{13 \xi ^3}{20}-\frac{1837 \xi
   ^2}{1800}+\frac{158 \xi
   }{225}-\frac{2291}{12600}\right)\nonumber
\eea

\bea
A^{(1)}_{(1,2)} & = & -\frac{659 \xi ^7}{176400}+\frac{367 \xi
   ^6}{9450}-\frac{143 \xi ^5}{700}+\frac{14873 \xi
   ^4}{25200}-\frac{14521 \xi ^3}{15120}+\frac{1759 \xi
   ^2}{2100}-\frac{536 \xi }{1575}+\frac{7421}{176400}\nonumber\\ 
& & +L_\xi
   \left(-\frac{\xi ^7}{280}+\frac{19 \xi
   ^6}{450}-\frac{37 \xi ^5}{225}+\frac{2 \xi
   ^4}{9}+\frac{9 \xi ^3}{40}-\frac{89 \xi
   ^2}{90}+\frac{907 \xi
   }{900}-\frac{536}{1575}\right)\nonumber
\eea

\bea
A^{(1)}_{(3,0)} & = & -\frac{311 \xi ^7}{66150}+\frac{11 \xi ^6}{252}-\frac{689
   \xi ^5}{3150}+\frac{10117 \xi ^4}{18900}-\frac{6259
   \xi ^3}{9450}+\frac{2561 \xi ^2}{6300}-\frac{323 \xi
   }{3150}+\frac{377}{132300} \nonumber\\ 
& & +L_\xi \left(-\frac{\xi ^7}{525}+\frac{\xi
   ^6}{50}-\frac{\xi ^5}{75}-\frac{19 \xi
   ^4}{90}+\frac{29 \xi ^3}{45}-\frac{119 \xi
   ^2}{150}+\frac{103 \xi
   }{225}-\frac{323}{3150}\right)\nonumber
\eea


\begin{thebibliography}{99}
\bibitem{babarspectrum}
B.~Aubert {\it et al.}  [BABAR Coll.],
arXiv:hep-ex/0408068.

\bibitem{fits}
B.~Aubert {\it et al.}  [BABAR Coll.],
Phys.\ Rev.\ Lett.\  {\bf 93} (2004) 011803
[hep-ex/0404017];
C.~W.~Bauer {\it et al.}, 
Phys.\ Rev.\ D {\bf 70} (2004) 094017
[hep-ph/0408002] v3;
O.~Buchmueller and H.~Flaecher, contribution to  the  CKM-2005 Workshop,
San Diego, march 2005, 
{\tt http://ckm2005.ucsd.edu/WG/WG2/thu3/henning-WG2-S2.pdf} and
arXiv:hep-ph/0507253.

\bibitem{btoc}
P.~Gambino and N.~Uraltsev,
Eur.\ Phys.\ J.\ C {\bf 34} (2004) 181
[arXiv:hep-ph/0401063].


\bibitem{Benson:2004sg}
D.~Benson, I.~I.~Bigi and N.~Uraltsev,
Nucl.\ Phys.\ B {\bf 710} (2005) 371
[arXiv:hep-ph/0410080].


\bibitem{btocpert}
  V.~Aquila, P.~Gambino, G.~Ridolfi and N.~Uraltsev,
  arXiv:hep-ph/0503083.




\bibitem{dfn}
F.~De Fazio and M.~Neubert,
  JHEP {\bf 9906}, 017 (1999)
  [arXiv:hep-ph/9905351].

\bibitem{1mb2}I.\,Bigi, N.\,Uraltsev and A.\,Vainshtein, 
{\it Phys.~Lett.}\ {\bf B293} (1992) 430 and
{\it Phys.\,Rev.\,Lett.}\ {\bf 71} (1993) 496;
B.\,Blok, L.\,Koyrakh, M.\,Shifman and A.\,Vainshtein,
{\it Phys.\,Rev.}\ {\bf D49} (1994) 3356;
A.~V.~Manohar and M.~B.~Wise,
Phys.\ Rev.\ D {\bf 49} (1994) 1310.

\bibitem{1mb3}
M.~Gremm and A.~Kapustin,  {\it Phys.\ Rev.}\ {\bf D55} (1997) 6924.

\bibitem{bds}
B.~Blok, R.~D.~Dikeman and M.~A.~Shifman,
Phys.\ Rev.\ D {\bf 51} (1995) 6167
[arXiv:hep-ph/9410293].

\bibitem{ritbergen}
T.~van Ritbergen,
Phys.\ Lett.\ B {\bf 454} (1999) 353
[arXiv:hep-ph/9903226];
M.~E.~Luke, M.~J.~Savage and M.~B.~Wise,
Phys.\ Lett.\ B {\bf 343} (1995) 329
[arXiv:hep-ph/9409287];
 P.~Ball, M.~Beneke and V.~M.~Braun,
 Phys.\ Rev.\ D {\bf 52} (1995) 3929 
[arXiv:hep-ph/9503492].

\bibitem{falkluke}
A.~F.~Falk and M.~E.~Luke,
Phys.\ Rev.\ D {\bf 57} (1998) 424
[arXiv:hep-ph/9708327].

\bibitem{kinetic}
I.~I.~Y.~Bigi, M.~A.~Shifman, N.~Uraltsev and A.~I.~Vainshtein,
Phys.\ Rev.\ D {\bf 56} (1997) 4017
[arXiv:hep-ph/9704245]
and 
Phys.\ Rev.\ D {\bf 52} (1995) 196
[arXiv:hep-ph/9405410].



\bibitem{imprecated}
D.~Benson, I.~I.~Bigi, T.~Mannel and N.~Uraltsev,
Nucl.\ Phys.\ B {\bf 665} (2003) 367
[arXiv:hep-ph/0302262].

\bibitem{q2spectrum}
A.~Czarnecki and K.~Melnikov,
Phys.\ Rev.\ Lett.\  {\bf 88} (2002) 131801
[arXiv:hep-ph/0112264].


\bibitem{tackmann}
M.~Battaglia, talk at BaBar $V_{ub}$ workshop, SLAC, december 2004;\\
K.~Tackmann, talk at CKM 2005, San Diego, march 2005, 
{\tt http://ckm2005.ucsd.edu/WG/WG2/fri2/tackman1-WG2-S4.pdf}.


\bibitem{WA}
  I.~I.~Y.~Bigi and N.~G.~Uraltsev,
  Nucl.\ Phys.\ B {\bf 423} (1994) 33
  [arXiv:hep-ph/9310285].

\bibitem{voloshin}
M.~B.~Voloshin,
Phys.\ Lett.\ B {\bf 515} (2001) 74
[arXiv:hep-ph/0106040].

\bibitem{fls}
A.~F.~Falk, M.~E.~Luke and M.~J.~Savage,
Phys.\ Rev.\ D {\bf 53} (1996) 2491
[arXiv:hep-ph/9507284].

\end{thebibliography}
\end{document}